\documentclass[preprint]{elsarticle}
\usepackage{lineno,hyperref}

\usepackage[linesnumbered,ruled]{algorithm2e}
\usepackage{vdrcmac_wo_fig_folder}
\usepackage{url}
\usepackage{mathtools}
\usepackage[dvipsnames]{xcolor}

\journal{Information Fusion}

\begin{document}

\begin{frontmatter}

\title{Fused Segmentation of Geometric Models for Myocardium and Coronary Artery via Medial Axis}

\author[myaddress1]{Jehyun Cha\fnref{myfootnote1}}
\ead{jhcha.vdrc@gmail.com}

\author[myaddress2]{Joonghyun Ryu\fnref{myfootnote1}}
\ead{jhryu@hanyang.ac.kr}

\author[myaddress3,myaddress4]{Jin-Ho Choi\corref{mycorrespondingauthor}\fnref{myfootnote2}}
\ead{jhchoimd@gmail.com}

\author[myaddress1, myadress2]{Deok-Soo Kim\corref{mycorrespondingauthor}\fnref{myfootnote2}}
\ead{dskim@hanyang.ac.kr}

\address[myaddress1]{School of Mechanical Engineering, Hanyang University, Korea}
\address[myaddress2]{Molecular Geometry and Vorononi Diagram Research Center, Hanyang University, Korea}
\address[myaddress3]{School of Medicine, Sungkyunkwan University, Korea}
\address[myaddress4]{Department of Emergency Medicine, Samsung Medical Center, Korea}
\fntext[myfootnote1]{Jehyun Cha and Joonghyun Ryu equally contributed.}
\fntext[myfootnote2]{Jin-Ho Choi and Deok-Soo Kim are co-corresponding authors.}

\begin{abstract}
Coronary arteries and their branches supply blood to myocardium.
The obstruction of coronary arteries results in significant loss of myocardium, called acute myocardial infarction, and the number one cause of death globally.
Hence, quantification of the regional amount of heart muscle subtended by obstructed coronary arteries is of critical value in clinical medicine.
However, the conventional methods are inaccurate and frequently disagree with clinical practices.
This study proposes a novel algorithm to segment regional myocardium-at-risk subtended by any potentially obstructed coronary artery.
Assuming the geometric triangular mesh models of coronary artery and myocardium derived from an individual cardiac computed tomography image, the proposed algorithm performs
(i) computation of the medial axis of the coronary artery and
(ii) segmentation of the coronary artery and myocardium using the medial axis.
The algorithm provides the fused segmentation of coronary artery and myocardium via the medial axis.
The computed result provides a robust mathematical linkage between myocardium-at-risk and supplying coronary arteries so that ischemic myocardial regions can be accurately identified, and
both the extent and severity of myocardial ischemia can be quantified effectively and efficiently.
Furthermore, the correspondence between segmented coronary artery and myocardium can be more importantly used for building optimization models of cardiac systems for various applications.
We believe that the proposed algorithm and implemented \texttt{VoroHeart} program, which is freely available at http://voronoi.hanyang.ac.kr/software/voroheart, will be an invaluable tool for patient-specific risk predictions and the treatment of obstructed coronary artery disease in clinical medicine.
The algorithmic accuracy and efficiency are theoretically asserted and experimentally verified.
\end{abstract}


\end{frontmatter}


\section{Introduction}
\label{sec:introduction}

Coronary artery disease is the number one cause of death worldwide and entrusts a huge socio-economic burden on the nations.
World Health Organization 2012 statistics reported that 7.4 million people died from coronary artery disease every year~\cite{WHO12}.
Atherosclerotic obstruction of coronary artery (CA) leads to loss of oxygenated blood supply to regional myocardium (heart muscle), which causes severe myocardial ischemia and acute myocardial infarction, i.e., a heart attack.

Localizing and assessing the extent of regional myocardium-at-risk subtended by obstructed CA is of critical importance in the diagnosis and decision of treatment~\cite{BruyneEtal14}.
A 17-piece myocardial model based on two-dimensional images is currently used by clinical guidelines as recommended by the American Heart Association~\cite{CerqueiraEtal02}\footnote{The \emph{17-segment} model popular in medicinal community is not used here to avoid the confusion with the \emph{segmentation} of data.}.
However, the model does not reflect individual structural variation of both CA and myocardium and frequently produces inaccurate assignments or disagreements between supplying CA and regional myocardium receiving blood~\cite{OrtizperezEtal08,JavadiEtal10}.
The establishment of an accurate and robust linkage between myocardial territory and supplying CA is required for the optimal diagnosis and treatment of CA disease~\cite{BruyneEtal12}.

Assuming that the triangular mesh model representation derived from an individual cardiac computed tomography image is available, this study presents a method to segment a myocardial 3D geometric model of a triangular mesh so that ischemic myocardial region can be accurately identified for each individual patient and both the extent and severity of myocardial ischemia can be quantified effectively and efficiently.
Nowadays, many programs are frequently used to extract the geometric model of a triangular mesh from computed tomography images~\cite{VitreaHome,AquariusHome,IntelliSpacePortalHome}.
There are two driving forces, which will be detailed in Sec.~\ref{subsec:experimental_results}, for the proposed research:
\begin{itemize}
\item Model Quantification for Geometric Analyses: A geometric model is convenient to quantify volume, boundary surface area, etc. which are of importance in clinical medicine~\cite{SaitoEtal05,SumitsujiEtal16,KurataEtal15,FrangiEtal01}.
\item Automatic Formulation for Optimization Problems: Given the quantification of a geometric model, optimization models can be formulated for automated decision-making in clinical practice. Cardiac stem cell therapy is such an example~\cite{Oettgen06,SegersLee08,ShafiqEtal16}.
\end{itemize}

\begin{sloppypar}
Figure~\ref{fig:computational_flow} shows the overview of the approach taken by the proposed research.
Given a cardiac computed tomography (cardiac CT) in the left of the figure, both CA and myocardium are derived and represented by three-dimensional geometric models of a triangular mesh~\cite{LorenzBerg06,PutterEtal06}.
Given 3D geometric models in the yellow box, the method computes the medial axis of the coronary arteries in the red circle and then performs the fused segmentation of the coronary arteries and myocardium in the blue box using the medial axis.
The \texttt{VoroHeart} program, which implemented the proposed algorithm, is freely available at http://voronoi.hanyang.ac.kr/software/voroheart.
Readers are recommended to refer to the demo video for the details of \texttt{VoroHeart}'s functions.
\end{sloppypar}

\FigOne{computational_flow}{0.8}
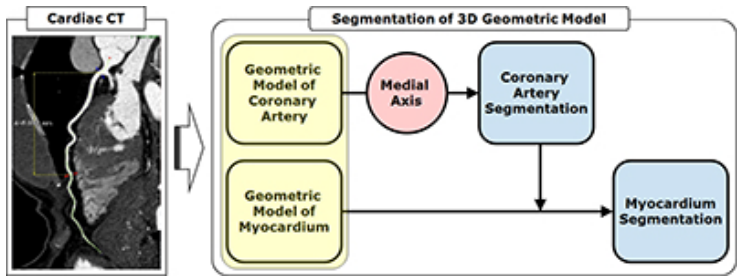
{Computational flow of this study.
The individual cardiac computed tomography (on the left) is reduced to 3D geometric models of a triangular mesh for myocardium and coronary arteries (in the yellow box).
Then, the proposed method computes the medial axis of the coronary arteries (in the red circle) and segments the coronary arteries and myocardium using the medial axis (in the blue box).}

Figure~\ref{fig:decomposition-of-heart-structure}(a) shows a cardiac CT image from which the three-dimensional geometric mesh models of Figs.~\ref{fig:decomposition-of-heart-structure}(b) through (f) are extracted.
The human heart consists of four chambers (i.e., two atriums and two ventricles), valves, CA, and proximal ascending aorta.
The entire heart structure is surrounded by pericardial fat (PF) as shown in Fig.~\ref{fig:decomposition-of-heart-structure}(b).
Figure~\ref{fig:decomposition-of-heart-structure}(c) shows CA, ascending aorta, left ventricle (LV), right ventricle (RV), and left atrium (LA) after the PF, right atrium, and pulmonary artery are removed from the heart structure.
Figure~\ref{fig:decomposition-of-heart-structure}(d) shows LV, aorta, and CA that consists of left CA (LCA) and right CA (RCA), both connected to the aorta.
Figure~\ref{fig:decomposition-of-heart-structure}(e) shows CA and LV which play a key role in cardiac function.
Figure~\ref{fig:decomposition-of-heart-structure} (f) show LV from a different view.

The statistics of the heart model in Fig.~\ref{fig:decomposition-of-heart-structure}, which is obtained from a teaching university hospital in Korea, are as follows.
Let $CA$ and $LV$ be the geometric models of a triangular mesh for the coronary artery and left ventricle, respectively.
Similarly, let $RV$, $LA$, and $PF$ be the mesh models for right ventricle, left atrium, and pericardial fat, respectively.
Let $|V(X)|$ and $|F(X)|$ denote the number of vertices and faces of $X$, respectively.
Then $|V(CA)|=14,990$,
$|F(CA)|=29,972$,
$|V(LV)|=34,642$,
$|F(LV)|=69,300$,
$|V(RV)|=20,262$,
$|F(RV)|=40,522$,
$|V(LA)|=9,950$,
$|F(LA)|=19,882$,
$|V(PF)|=99,327$, and
$|F(PF)|=199,600$.

\FigSix{decomposition-of-heart-structure}{0.2}{0.288}{0.288}{0.288}{0.288}{0.288}
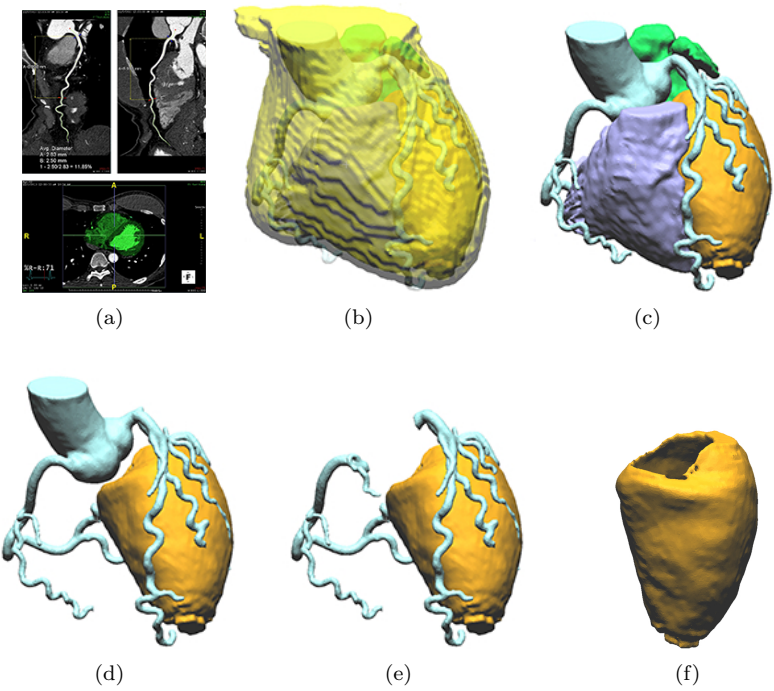
{Heart structure.
(a) Cardiac computed tomography (cardiac CT),
(b) the entire heart surrounded by pericardial fat (PF),
(c) coronary artery (CA), proximal ascending aorta (sky blue), left ventricle (LV) (dark yellow), right ventricle (RV) (light purple), and left atrium (LA) (green) after the PF, right atrium, and pulmonary artery are removed,
(d) LV, ascending aorta, and CA,
(e) CA and LV which play a key role for cardiac function, and
(f) LV viewed from a different orientation}

We discuss a geometric method to segment myocardial region of the left ventricle (LV) into subregions, where each corresponds to a coronary artery CA piece and/or a concatenation of consecutive downhill pieces of CA.
In this study, CA is modeled as a closed 2-manifold triangular mesh of a single shell with neither a mesh boundary nor a handle.
This implies that CA is represented by a mesh surface with neither wall thickness nor any void.
LV is also similarly modeled, yet it is interpreted to possess a thickness corresponding to the heart wall, which should be inferred from the distance between the appropriate triangular faces in the neighborhood (See Fig.~\ref{fig:decomposition-of-heart-structure}(f)).

The remainder of this paper is organized as follows:
Section~\ref{sec:literature-review} reviews related studies.
Section~\ref{sec:preprocessing-for-coronary-artery-medial-axis-computation} presents an algorithm for extracting an adjacency tree from an adjacency graph, which is constructed from a constrained Delaunay triangulation of the coronary artery.
Section~\ref{sec:construction-of-medial-axis-from-V-tree} presents the medial axis computation by refining the adjacency tree.
Section~\ref{sec:segmentation-and-medical-applications} presents the segmentation of the left ventricle and coronary artery.
Section~\ref{sec:experiments-and-discussion} presents the algorithm summary and experimental result.
Section~\ref{sec:conclusion} concludes the paper.

\section{Literature review}
\label{sec:literature-review}

This study discusses the segmentation of 3D geometric mesh models derived from cardiac CT images for the left ventricle and coronary artery using the medial axis of the coronary artery.
The paper discusses two technical issues:
(i) the medial axis of coronary artery and
(ii) the segmentation of coronary artery and left ventricle using the medial axis that establishes their correspondence.
We thus review these two issues, i.e., medial axis computation and the segmentation of image or mesh.

\begin{sloppypar}
The medial axis, sometimes called the symmetric axis or skeleton \cite{Kirkpatrick79},
was first introduced by Blum in 1967 in order to describe biological shapes~\cite{Blum67} and was extensively used for diverse applications
such as shape description/matching~\cite{BlumNagel78,NackmanPizer85,PizerEtal87},
surface reconstruction~\cite{AmentaEtal98,JalbaEtal13},
animation~\cite{WadeParent02},
smoothing or sharpening of shape~\cite{HoDyer86},
motion planning~\cite{HollemanKavraki00}, and
mesh generation~\cite{TamArmstrong91,LinardakisChrisochoides08}.
\end{sloppypar}

Voronoi diagram (VD) and its dual Delaunay triangulation (DT) is one of the most fundamental tool for shape analysis and spatial reasoning.
VD and DT has many applications such as
coverage maximization of wireless sensor network (WSN)~\cite{Abo-ZahhadEtal16},
deployment schemes for sensor coverage in WSN~\cite{SenouciEtal15},
analysis of data collected from WSN~\cite{BosmanEtal17}, and
facility location~\cite{PengEtal14}, etc.
It is known that a medial axis for a planar shape can be correctly and efficiently computed using the Voronoi diagram of a simple polygon~\cite{Kirkpatrick79,Lee82} which can also be used for the offset computation of the polygon by representing the Voronoi edges with rational quadratic {B}$\acute{e}$zier curves~\cite{Kds98,KdsEtal95}.
On the other hand, its counterpart in 3D, called a medial surface, may contain both surface patches as well as degenerating curves ~\cite{JalbaEtal13,GongBertrand90}, thus leaving its computation a challenge.
Culver et al. presented an algorithm and its implementation for polyhedra using exact arithmetic~\cite{CulverEtal04}.
However, the algorithm turned out impractical due to the enormous computational requirement to compute the correct medial axis with polyhedra of even a moderate size of hundreds of faces due to the algebraic complexity of the medial axis.
Another important issue related with the exact computation approach is that a medial axis can have many insignificant parts related with a tiny disturbance of polyhedron geometry because the medial axis is very sensitive to perturbation~\cite{ShakedBruckstein98,ChoiSeidel04}.
Therefore, an approximation approach is sufficiently justified.

An obvious approach to the approximation of a medial axis might be first to compute the Voronoi diagram of points on a model boundary and then to remove the insignificant parts of the Voronoi diagram~\cite{AttaliMontanvert97,AmentaEtal01,DeyZhao04}.
This approach seems reasonable as Brandt showed in 2D that the Voronoi vertices inside the shape boundary converges to its medial axis as the sampling rate increases~\cite{Brandt94}.
However, practical consideration of the trade-off between the computational requirement and solution quality due to the number of sampling points on the model boundary becomes a major bottleneck of this approach.
Attali and Montanvert proposed an approximation algorithm of a 3D medial axis using the Voronoi diagram of intersection points of 3D spherical balls, which approximates the shape~\cite{AttaliMontanvert97}.

Based on the research experience of both segmentation of images~\cite{JamesDrsarathy14,MakniEtal14,Mignotte14,SchwarzKasparek14,SankaranEtal17} and segmentation of mesh models~\cite{Shamir08a,AuEtal12,ZhengEtal12a}, the segmentation of myocardium and blood vessel was intensively studied.
While several image-based studies were reported~\cite{Suri00,MitchellEtal01,Paragios03,LesageEtal09,KurataEtal15,CabriaGondra17,TangEtal16}, one noteworthy work was to segment the computed tomography image of the left ventricle LV using that of the coronary artery CA, where a user manually picks some voxel points belonging to CA so that each voxel of LV can be assigned to its closest picked point~\cite{KurataEtal15,DebarbaEtal10}.
An application of the medial axis based on an image-based approach was reported for analyzing the morphometry, such as diameters and branching pattern of CA~\cite{WischgollEtal08}.

An improvement was made to take advantage of the mesh representation of geometric modeling.
For the segmentation of myocardial mesh in 3D, a notably improved algorithm was reported that combined two representations~\cite{TermeerEtal10}, i.e., image and mesh:
i) For CA, tomographic image representation - The centerline was computed via an image processing technique including the identification of branch points~\cite{LorenzEtal03}, which was then projected to the surface of LV mesh;
ii) For LV, mesh representation - The Voronoi diagram of the projected points on the LV surface was computed with the geodesic distance metric so that the collection of the Voronoi cells belonging to the projection of the voxels of a particular artery piece could provide the segmentation information of LV.

\section{Extraction of the adjacency tree for the coronary artery}
\label{sec:preprocessing-for-coronary-artery-medial-axis-computation}

This section explains an algorithm to extract an adjacency tree from an adjacency graph, which is constructed using a constrained Delaunay triangulation of the coronary artery $CA$.
We will transform the adjacency tree to the medial axis of the $CA$, which is a one-dimensional curve-skeleton~\cite{LiuEtal10a,CorneaEtal07}.

\subsection{Constructing the adjacency graph from constrained Delaunay triangulation}
\label{subsec:CDT-and-adjacency-graph}

The medial axis of a simple polygon in the plane is a subset of the Voronoi diagram of the polygon~\cite{Kirkpatrick79,Lee82}.
Figures~\ref{fig:MA-CDT-VG-VD}(a) and (b) show the correct medial axis of a simple polygon and the (interior) Voronoi diagram of the polygon, respectively.
Consider a reflex vertex $v$ in Fig.~\ref{fig:MA-CDT-VG-VD}(b) where the internal angle between the incident edges is greater than 180 degrees.
Removing Voronoi edges incident to all reflex vertices reduces the Voronoi diagram to the correct medial axis~\cite{Lee82,KdsEtal95}.

However, this idea cannot be directly applied to the three-dimensional counterpart because the correct Voronoi diagram of a polyhedron is difficult to compute in general~\cite{CulverEtal04}.
Instead, an obvious and direct approach might be to use the ordinary Voronoi diagram of sampling points on the shape boundary, as this type of Voronoi diagram can be easily computed.
Then, a medial axis approximation can be obtained by a postprocessing of pruning the insignificant substructure of the Voronoi diagram.

The proposed algorithm is based on the constrained Delaunay triangulation (CDT) which is the dual structure of the constrained Voronoi diagram (Note that the Delaunay triangulation is the dual of the ordinary Voronoi diagram of points)~\cite{Aurenhammer91,OkabeEtal99}.
The idea is explained by the same figure.
Figure~\ref{fig:MA-CDT-VG-VD}(c) shows CDT for point generators on shape boundary where the point set contains some sampled points (i.e., the filled rectangles) in addition to the vertices of the polygon.
Figure~\ref{fig:MA-CDT-VG-VD}(d) shows the adjacency graph, which represents the adjacency among the triangles of CDT by connecting the centers of the circumcircles of triangles (See Definition~\ref{def:adjacency-graph}).
Note that three dotted circumcircles do not contain any other point generators, thus leading to the Delaunay property.
Figure~\ref{fig:MA-CDT-VG-VD}(e) shows the adjacency graph on the top of the medial axis.
Observe that the adjacency graph relatively well approximates the medial axis.
The main idea of this study for the three-dimensional medial axis starts from this simple yet important observation.

\FigFive{MA-CDT-VG-VD}{0.6}{0.6}{0.6}{0.6}{0.6}
{
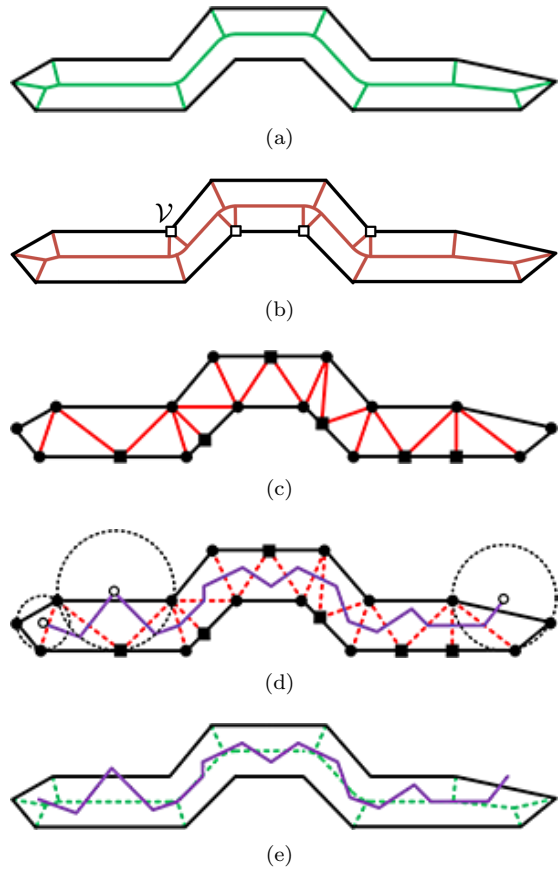
Various geometric constructs for a polygon.
(a) The medial axis,
(b) the interior Voronoi diagram of the polygon,
(c) the constrained Delaunay triangulation (CDT) of points on the polygon boundary,
(d) CDT and the corresponding adjacency graph, and
(e) both adjacency graph and medial axis.
}

The algorithm of CDT in three-dimensional space was reported~\cite{Shewchuk03,SiGartner11,SiShewchuk14} and its well-known implementation $TetGen$ is available~\cite{Si08a,Si13,Si15}.
Let $CDT = (V, E, F, C)$ be the CDT of the geometric mesh model $CA$ for coronary artery where $V$, $E$, $F$, and $C$ are the sets of vertices, edges, faces, and tetrahedral cells of the triangulation, respectively.
\begin{definition_} \emph{(\textbf{Adjacency graph} $\mathcal{G}$)}
\label{def:adjacency-graph}
Let $\mathcal{G} = (N, L) \equiv \mathcal{G}(N, L)$ be the adjacency graph (adj-graph) of $CA$ where $N$ and $L$ are the sets of nodes and links, respectively.
Then $n \in N$ is in one-to-one correspondence to $c \in C$, and the coordinate of the circumsphere center of $c$ becomes the attribute of $n$.
Two cells $c_i$, $c_j \in C$ are defined to be \emph{adjacent} to each other if $c_i$ and $c_j$ share a common triangular face.
Then each pair of adjacent cells in $C$ defines $l \in L$.
\end{definition_}
$\mathcal{G}$ has the following properties.
\begin{itemize}
  \item $\mathcal{G}$ may have cycles.
  \item $\mathcal{G}$ may have one or more nodes located outside the model boundary.
\end{itemize}
The construction of an adj-graph is obvious.
Given $CDT = (V, E, F, C)$, suppose that $c \in C$ has $d$ adjacent cells
(i.e., $c$ has $d$ neighboring cells where $d\in\{1, 2, 3, 4\}$ and each cell share one of its face with $c$).
If the triangulation is stored in the data structure, such as the simplicial complex data structure~\cite{KdsEtal06}, the traversal from a cell to its adjacent cell take $O(1)$ time in the worst case.
As there are $O(m)$ adjacency relationships for CDT with $m$ triangular faces, the construction of the adj-graph can be done in linear time with respect to the number of tetrahedral cells in $CDT$.

\subsection{Extracting the adjacency tree from the adjacency graph by removing cycles}
\label{subsec:transform-of-adjacency-graph-to-adjacency-tree}

As the overall topology of the coronary artery is a tree, the corresponding medial axis should also be a tree.
However, an adjacency graph \emph{adj-graph} may have cycles depending on the arrangement of cells in the constrained Delaunay triangulation.
We should remove the cycles to reduce an adj-graph to a tree.
\begin{definition_} \emph{(Adjacency tree $\mathcal{T}$)}
The adjacency tree (adj-tree) $\mathcal{T}(N^\mathcal{T}, L^\mathcal{T})$ corresponding to an adjacency graph (adj-graph) $\mathcal{G}(N, L)$ of $CA$ is a connected subgraph of $\mathcal{G}$ without any cycles where $N^\mathcal{T} \subseteq N$, $L^\mathcal{T} \subseteq L$.
\end{definition_}
The adj-tree may not be unique because there are in general multiple ways to remove cycles of a graph.
Computing the shortest path tree (SPT) of an adj-graph, for example using the Dijkstra algorithm~\cite{Dijkstra59}, might be a simple yet effective approach.
To launch the SPT algorithm, it is necessary to determine the root node from which the entire tree can be constructed.
While determination of the root node in reality requires information beyond geometry, an observation is that the shape of $CA$ becomes narrower as it approaches an end tip.
This implies that the root node tends to be most volumetric among the cells in $CA$.
Hence, in the automatic mode, we define the root node as the cell with the largest face in the entire triangulation.
We also have an additional mode to manually select the root node.

Figure~\ref{fig:root_dependency}(a) shows a schematic diagram for $CA$ and its ideal medial axis.
Assuming that the node in the northernmost tip is the root node $n_{root}$, the adj-tree can be extracted with the Dijkstra algorithm by finding the shortest path from $n_{root}$ to each node of the corresponding adj-graph.
However, the computed shortest path tree can be problematic in that it can be somewhat different from the ideal medial axis.
It turns out that the branching node of the shortest path tree moves closer to the root node.
Figures~\ref{fig:root_dependency}(b) and (c) show that this problem indeed occurs and the choice of another node as the root does not solve the problem, respectively.
As the tree can branch off at a node closer to the root than at the desirable node, this phenomenon is called a $premature branching$ of the shortest path tree.
In real $CA$ models, the premature branching can be of significance as shown in
Fig.~\ref{fig:premature_bifurcation}.
The zoom-up shows that the tree branches off at a node much closer to the root node than at the node where it actually should.

\FigThree{root_dependency}{0.25}{0.25}{0.25}
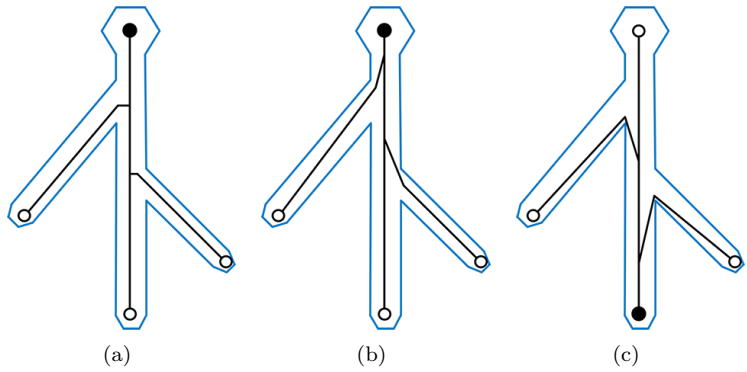
{The root dependency of the shortest path tree construction.
(a) The ideal medial axis of a model,
(b) the shortest path tree computed with the northernmost tip as the root node, and
(c) another shortest path tree with the southernmost tip as the root node.
}

\FigOne{premature_bifurcation}{0.7}
{
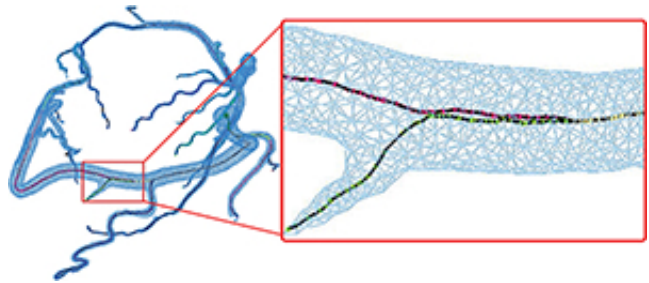
The premature branching of a shortest path tree.
}

Let the shortest path tree above be the forward shortest path tree $\mathcal{T}^{SPT}_{FWD}$ computed by the forward pass of Dijkstra.
Each shortest path is called the forward shortest path.
In order to avoid premature branching, we modify the forward shortest path tree $\mathcal{T}^{SPT}_{FWD}$ as follows.
Suppose that there are $m$ leaf nodes in $\mathcal{T}^{SPT}_{FWD}$ and thus $k$ paths from the root node.
We store the paths in the priority queue $Q$ according to the path length in non-increasing order (i.e., the root of $Q$ corresponds to the longest path in $\mathcal{T}^{SPT}_{FWD}$).
Each node of $Q$ also stores the leaf node of each path of $\mathcal{T}^{SPT}_{FWD}$.

Figure~\ref{fig:reverse-growing-algorithm}(a) shows $\mathcal{T}^{SPT}_{FWD}$ for $k = 9$ where the root (black filled circle) and leaf  (unfilled circle) nodes are shown.
Let $\pi_1^{FWD}$ be the forward shortest path corresponding to the root of $Q$, which is the longest path of $\mathcal{T}^{SPT}_{FWD}$ from the root node $n_{root}$ to a leaf node, say $n_1$.
Let $\mathcal{T}_1 = (N^\mathcal{T}_1, L^\mathcal{T}_1)$ be the initial adj-tree, where $N^\mathcal{T}_1$ and $L^\mathcal{T}_1$ denote the sets of all nodes and all links in $\pi_1^{FWD}$, respectively.
Figure~\ref{fig:reverse-growing-algorithm}(b) shows $\mathcal{T}_1$.
For the next iteration, $Q$ is updated by removing its current root.

Consider $\pi_2^{FWD}$ corresponding to the next root of $Q$, which is the second longest path of $\mathcal{T}^{SPT}_{FWD}$ from $n_{root}$ to another leaf node, say $n_2$.
Then we compute the shortest path $\pi_2^{BWD}$ of adj-graph $\mathcal{G}$ from $n_2$ to $\mathcal{T}_1$ by applying the Dijkstra algorithm in a backward fashion.
The topological distance from a node to a tree is stated in Definition~\ref{def:geodesic-from-node-to-tree} below.
We grow $\mathcal{T}_1$ to $\mathcal{T}_2$ by concatenating $\pi_2^{BWD}$ to $\mathcal{T}_1$ (Definition~\ref{def:concatenation-tree-and-path} below states the concatenation of a tree and a path).
Figure~\ref{fig:reverse-growing-algorithm}(c) illustrates the construction of $\mathcal{T}_2$.

The next root of $Q$ corresponds to the third longest path, say $\pi_3^{FWD}$ from $n_{root}$ to the leaf node, say $n_3$.
Then we compute the shortest path $\pi_3^{BWD}$ from $n_3$ to $\mathcal{T}_2$ and concatenate it to $\mathcal{T}_2$ to produce $\mathcal{T}_3$.
Figure~\ref{fig:reverse-growing-algorithm}(d) illustrates $\mathcal{T}_3$.
We repeat this process while the priority queue is non-empty (See Figures~\ref{fig:reverse-growing-algorithm}(e) and \ref{fig:reverse-growing-algorithm}(f)).
The above each shortest path $\pi^{BWD}$ is called the backward shortest path and
each shortest path tree is called the backward shortest path tree computed by the backward pass of Dijkstra.
Figure~\ref{fig:reverse-growing-algorithm}(f) shows the final backward shortest path tree computed from Fig.~\ref{fig:reverse-growing-algorithm}(a).
\begin{definition_}  \emph{(Distance between a node and a subtree)}
\label{def:geodesic-from-node-to-tree}
Suppose $\mathcal{T}(N^\mathcal{T},L^\mathcal{T}) \subseteq G(N,L)$.
The topological distance between a node $n \in N$ and a tree $\mathcal{T}$ is given as
\begin{equation}
\label{eqn:distance-between-tree-and-node}
dist(n, \mathcal{T}) = \min_{n_i \in N^\mathcal{T}} d(n, n_i)
\end{equation}
where $d(n,n_i)$ is the topological distance between $n$ and an arbitrary node $n_i \in N^\mathcal{T}$ through the shortest path between them.
\end{definition_}
\begin{definition_}  \emph{(Concatenation)}
\label{def:concatenation-tree-and-path}
Given a tree $\mathcal{T}(N^\mathcal{T},L^\mathcal{T})$ and a path $\pi \subseteq G$,
the \emph{concatenation} $\mathcal{T} \bigoplus \pi$ is defined as the addition of the nodes and links of $\pi$ to $N^\mathcal{T}$ and $L^\mathcal{T}$, respectively.
\end{definition_}
\begin{definition_} \emph{(Intermediate adj-tree)}
Each subtree $\mathcal{T}_i \subseteq G(N,L)$ defined by the concatenations of a set of paths is an \emph{intermediate adj-tree}.
\end{definition_}
Note that Eq.~(\ref{eqn:distance-between-tree-and-node}) prevents premature branching because the distance is now defined from each leaf node to the intermediate adj-tree rather than from the root node.

Let $\pi_i^{BWD}$ be a backward shortest path from the leaf node $n_i$ of adj-graph $\mathcal{G}$, which corresponds to the current root of $Q$ at the $i$-th step.
Then the following lemma holds.
\begin{lemma_} \emph{(Construction of adj-tree by concatenation)}
\label{lemma:construction-of-intermediate-adj-tree}
\begin{small}
\begin{eqnarray}
\label{eqn:concatenation-between-tree-and-node}
\mathcal{T}_n &=& \mathcal{T}_{n-1} \bigoplus \pi_n^{BWD} = \mathcal{T}_{n-2} \bigoplus \pi_{n-1}^{BWD} \bigoplus \pi_n^{BWD} = \ldots \nonumber \\
= &\mathcal{T}_1& \bigoplus \pi_2^{BWD} \bigoplus \pi_3^{BWD} \ldots \bigoplus \pi_{n-1}^{BWD} \bigoplus \pi_n^{BWD}
\end{eqnarray}
\end{small}
\end{lemma_}
Be aware that $\mathcal{T}_n$ can be different from $\mathcal{T}^{SPT}_{FWD}$ and the processes in the following sections are based on $\mathcal{T}_n$.

\FigSix{reverse-growing-algorithm}{0.3}{0.3}{0.3}{0.3}{0.3}{0.3}
{
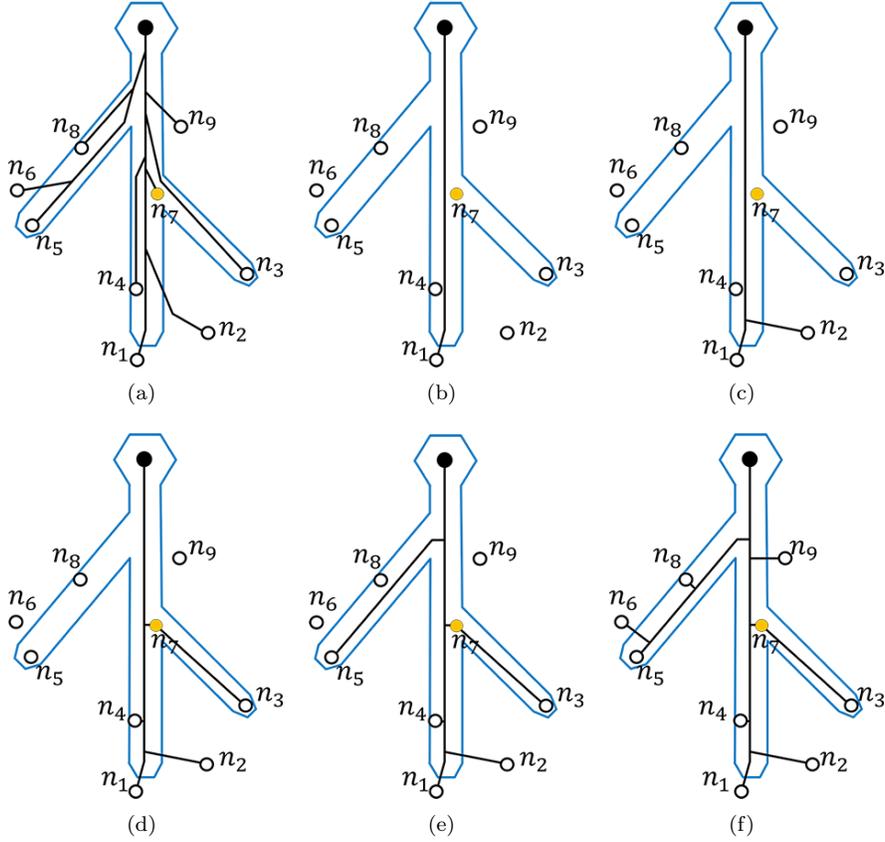
The extraction of an adjacency tree from an adjacency graph by forward and backward shortest paths.
(a) Forward shortest path tree $\mathcal{T}^{SPT}_{FWD}$ of an adjacency graph from the northernmost root node.
The subtrees are updated by concatenating the backward shortest path from the current leaf node to the previous subtree as follows:
(b) by concatenating $\pi_1^{BWD}$ from $n_1$ (the initial adjacency tree),
(c) by concatenating $\pi_2^{BWD}$ from $n_2$,
(d) by concatenating $\pi_3^{BWD}$ and $\pi_4^{BWD}$ from $n_3$ and $n_4$, respectively,
(e) by concatenating $\pi_5^{BWD}$ from $n_5$, and
(f) by concatenating $\pi_6^{BWD}$, $\pi_8^{BWD}$, $\pi_9^{BWD}$ from $n_6$, $n_8$, $n_9$, respectively (the final adjacency tree).
}

\begin{algorithm}[htpb]
\SetAlgoLined

\SetKwInOut{Input}{input}\SetKwInOut{Output}{output}
\Input{adjacency graph $\mathcal{G}$, root node $n_{root}$}
\Output{adjacency tree $\mathcal{T}$}
\BlankLine

construct the forward shortest path tree $\mathcal{T}^{SPT}_{FWD}$ of $\mathcal{G}$ with $n_{root}$\;

\For{a leaf node $n_i$ in $\mathcal{T}^{SPT}_{FWD}$}
{
    compute a path $\pi_i^{FWD}$ from $n_{root}$ to $n_i$\;
    compute the path length $|\pi_i^{FWD}|$ of $\pi_i^{FWD}$\;
    push $n_i$ and $\pi_i^{FWD}$ into the priority queue $Q$ according to the non-increasing order of $|\pi_i^{FWD}|$\;
}

construct the initial adjacency tree $\mathcal{T}_1$\;
\While{$Q$ is not empty}
{
    pop the root $n^Q_i$ of $Q$\;
    compute the backward shortest path $\pi_i^{BWD}$ from a leaf node $n_i$ to $\mathcal{T}_{i-1}$
    where $n_i$ corresponds to $n^Q_i$\;
    \If{the path length $|\pi_i^{BWD}| > 0$}
    {
        concatenate ($\mathcal{T}_{i-1} \bigoplus \pi_i^{BWD}$)\;
    }
}
\caption{Extracting Adjacency Tree}\label{algo:construct-adjacency-tree}
\end{algorithm}

The proposed tree extraction algorithm is summarized in Algorithm~\ref{algo:construct-adjacency-tree}.
The code chunk line 1 through 6 computes the forward shortest path tree $\mathcal{T}^{SPT}_{FWD}$ and stores both the leaf nodes and the corresponding paths in the priority queue.
Line 7 constructs the initial adj-tree $\mathcal{T}_1$.
Lines 8 through 14 iteratively update the subtree by concatenating a backward shortest path from a leaf node to the previous intermediate adj-tree.
Note that line 11 implies that the length of a backward path $\pi^{BWD}$ can be zero when the leaf node for $\pi^{BWD}$ is already included in any previously updated subtree.
For example, a leaf node $n_7$ is included in the backward shortest path $\pi_3^{BWD}$ as shown in Fig.~\ref{fig:reverse-growing-algorithm}(d).
The length of the backward path $\pi_7^{BWD}$ from $n_7$ is zero, thus $\pi_7^{BWD}$ is not concatenated any more.

The Dijkstra algorithm takes $O(|L| + |N| \log |N|)$ time in the worst case if it is based on a priority queue implemented by a Fibonacci heap where $|N|$ and $|L|$ represent the numbers of nodes and links, respectively~\cite{FredmanTarjan84}.
Note that the Dijkstra algorithm can deteriorate to $O({|N|}^3)$ time in the worst case if it is implemented otherwise.
The proposed algorithm runs two passes of Dijkstra:
one for the forward pass and the other for the backward pass.
The forward pass computes the shortest paths of all nodes from the root, whereas the backward pass consists of the runs for computing the backward shortest path from each leaf node to an intermediate adj-tree.
Hence, the worst-case time complexity of Algorithm~\ref{algo:construct-adjacency-tree} becomes $O(|N||L|+{|N|}^2 log{|N|})$ using a Fibonacci heap.

\section{Transformation of adjacency tree to medial axis}
\label{sec:construction-of-medial-axis-from-V-tree}

The adjacency tree \emph{adj-tree} above becomes the initial representation of the medial axis, which is fine-tuned in this section.

\subsection{Removing outrageous nodes}
\label{subsubsec:removing-outrageous-nodes}

The medial axis of a shape should be located inside the shape boundary.
However, the adj-tree obtained in the previous section may have some extraneous nodes that are placed outside the geometric mesh model $CA$ of the coronary artery.
Figure~\ref{fig:outrageous-nodes} shows $CA$ and the corresponding adj-tree.
As shown in the close-up, the adj-tree has several blue leaf nodes emanating to the outside of $CA$.
This section presents an algorithm to remove those extraneous nodes.

\FigOne{outrageous-nodes}{0.6}
{
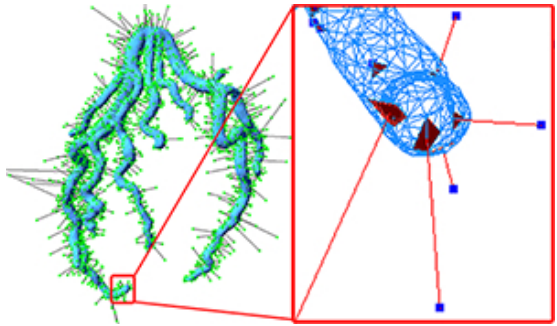
Example of outrageous nodes.
The left coronary artery $LCA$ and its adj-tree $\mathcal{T}$.
There are many outrageous nodes which are placed outside $LCA$.
}

\begin{definition_} \emph{(Outrageous node)}
Given an adj-tree $\mathcal{T}(N, L)$, if $n \in N$ is placed outside $CA$, $n$ is called an \emph{\textbf{outrageous node}} and the corresponding CDT cell is called an \emph{\textbf{outrageous cell}}.
\end{definition_}

It turns out that an outrageous cell tends to be relatively flat and is located near the boundary of $CA$~\cite{AmentaEtal01}.
Figure~\ref{fig:outrageous-nodes}, in the close-up, shows some (blue) outrageous nodes and the corresponding (red) flat cells near the boundary.
We will remove those nodes from $\mathcal{T}$ via two steps as follows:
(i) Identify the outrageous nodes and (ii) remove those nodes.

A simple yet effective way to identify the outrageous nodes is to check if the intersection between links and the boundary of $CA$.
Consider a link $l$ whose start node is inside $CA$.
The end node of the link $l$ will be inside if $l$ does not intersect $CA$ or has an even number of intersections.
On the other hand, the end node will be outside if $l$ has an odd number of intersections.
Mostly, the number of intersections will be less than two.
If we pick the links incident to the end node, this process can be repeated because we have always one visited node (a node whose status is known) and the other unvisited node for each link.
After all nodes are visited, all of the outrageous nodes are recognized.

After identifying the outrageous nodes, we traverse the adj-tree, starting from an outrageous leaf node, and collect the outrageous nodes until we cross the $CA$ boundary.
Then we remove the collected nodes safely.
The bucket system is exploited in order to localize the candidate faces of $CA$ for the intersection.
Assuming that the number of candidate faces for the intersection of each link is $O(1)$, the algorithm takes $O(|N|)$ time in the worst-case scenario where $|N|$ is the number of nodes of adj-tree.

\subsection{Shaving hairs}
\label{subsubsec:shaving-hairs}

The adj-tree after the removal step usually contains tiny subtrees that constitute an undesirable part of the medial axis.
Such a subtree is called a $hair$ because its contribution to a medial axis is negligible.
In this section, we discuss an operation to remove hairs, called $shaving$.
Figures~\ref{fig:example-of-hairs-in-3D}(a) and (b) show the adj-tree before and after shaving the (yellow) hairs, respectively.

\FigTwo{example-of-hairs-in-3D}{0.6}{0.309}
{
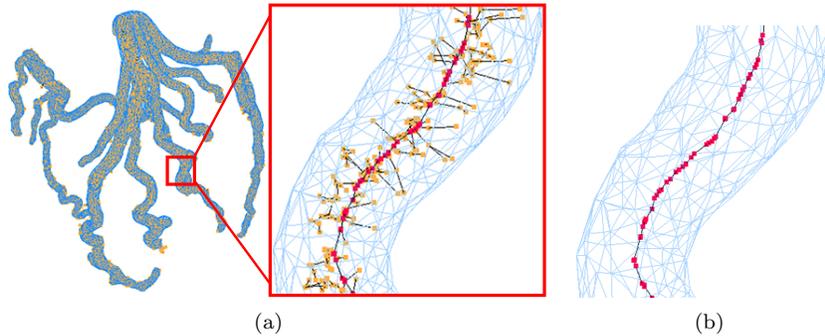
Example of hairs.
(a) Before and
(b) after the shaving operation.
}

We measure the contribution of each subtree, say $X$, of an adj-tree $\mathcal{T}$ to the corresponding medial axis as follows.
Given an intermediate tree $\mathcal{T}_{i-1}$, consider adding a subtree $X_i$ to $\mathcal{T}_{i-1}$ to produce $\mathcal{T}_i$.
If the contribution of $X_i$ is significant, it is $X_i$ is considered a new subtree of $\mathcal{T}_i$.
Otherwise, we consider it a hair.

\begin{definition_}
\emph{(Distance between trees)}
The distance between two trees $\mathcal{T}_i(N_i,L_i)$ and $\mathcal{T}(N,L)$ where $\mathcal{T}_i \subseteq \mathcal{T}$ is
\begin{equation}
\label{eqn:distance-of-abstract-tree}
\Delta_{i} = \Delta(\mathcal{T}_i,\mathcal{T}) = \sum_{n_j \in N} dist(n_j,\mathcal{T}_i)
\end{equation}
\begin{equation}
\label{eqn:distance-of-distances}
\delta_i = \Delta_{i-1} - \Delta_{i} = \Delta(\mathcal{T}_{i-1},\mathcal{T}) - \Delta(\mathcal{T}_i,\mathcal{T})
\end{equation}
where $dist(n_j,\mathcal{T})$ is the topological distance between $n_j$ and $\mathcal{T}$ as defined in Eq.~(\ref{eqn:distance-between-tree-and-node}).
\end{definition_}
Note that $\Delta_{i}=0$ if and only if $\mathcal{T}_i \equiv \mathcal{T}$.
If $\mathcal{T}_i \not \equiv \mathcal{T}$, $\Delta(\mathcal{T}_i,\mathcal{T}) > 0$.
Hence, $\Delta(\mathcal{T}_i,\mathcal{T})$ can be an indicator that tells us how close the intermediate adj-tree $\mathcal{T}_i$ is to $\mathcal{T}$.

\begin{lemma_} \emph{(Monotonicity of tree distance)} (For proof, see~\ref{appendix:proof-1}.)
\label{lemma:monotonicity-of-distance-of-abstract-tree}

\begin{equation}
\label{eqn:monotonicity-of-distance-of-abstract-tree}
\Delta_{i-1} > \Delta_{i}
\end{equation}
\end{lemma_}
Hence, $\Delta_i$ is non-increasing and $\delta_i > 0$.
Lemma~\ref{lemma:monotonicity-of-distance-of-abstract-tree} states that an intermediate adj-tree is closer to the original adj-tree as we concatenate more paths.
However, the marginal increase of the closeness of each incrementally concatenated path should be investigated since that of one path can be significantly different from that of another.
We evaluate $\delta_i$ for each root, say $n^Q_i$ in the $i$-th iteration, of a priority queue $Q$, which corresponds to a path $\pi_i$ in $\mathcal{T}$.
$\pi_i$ is concatenated to $\mathcal{T}_{i-1}$ if $\delta_i > \epsilon$, where $\epsilon$ is a parameter given by the user.

The proposed hair-shaving algorithm is summarized in Algorithm~\ref{algo:shaving-hairs}.
Lines 1 through 5 compute the path from the root node to each leaf node and store each path and its corresponding leaf node in the priority queue $Q$ according to a non-increasing order of path length.
Lines 6 through 15 compute the distance $\Delta(\mathcal{T}_i,\mathcal{T})$ of each intermediate adj-tree $\mathcal{T}_i$ which is constructed by concatenating each path.
Then $\delta_i$ of Eq.~(\ref{eqn:distance-of-distances}) is also computed.
Lines 12 through 14 incrementally construct the shaved adj-tree.
Note that line 13 concatenates a path $\pi_i$ to the current shaved tree only if the distance reduction $\delta_i$ induced by $\pi_i$ is larger than a given threshold $\epsilon$.
In this study, we have used the average distance reduction $\sum_{i=1}^n \frac{\delta_i} {n}$ of intermediate adj-trees for $\epsilon$.
Algorithm~\ref{algo:shaving-hairs} takes $O(|N|^2)$ time in the worst-case scenario, where $|N|$ is the number of nodes in the unshaved tree.

\begin{algorithm}[htpb]
\SetAlgoLined

\SetKwInOut{Input}{input}\SetKwInOut{Output}{output}
\Input{  unshaved adjacency tree $\mathcal{T}$}
\Output{ shaved adjacency tree $\mathcal{T}^{Shaved}$}
\BlankLine

\For{a leaf node $n$ in unshaved tree $\mathcal{T}$}
{
    compute a path $\pi$ from the root node $n_{root}$ to $n$\;
    compute the path length $|\pi|$ of $\pi$\;
    push $n$ and $\pi$ into the priority queue $Q$ according to the non-increasing order of $|\pi|$\;
}

construct the initial adjacency tree $\mathcal{T}_1$\;
initialize $\mathcal{T}^{Shaved} \equiv \emptyset$

\While{$Q$ is not empty}
{
    pop the root node $n^Q_i$ in $Q$\;
    construct $\mathcal{T}_i$ by the concatenation ($\mathcal{T}_{i-1} \bigoplus \pi_i$)
    where $\pi_i$ corresponds to $n^Q_i$\;
    compute $\Delta(\mathcal{T}_i,\mathcal{T})$\ and $\delta_i = \Delta(\mathcal{T}_{i-1},\mathcal{T}) - \Delta(\mathcal{T}_i,\mathcal{T})$\;
    \If{$\delta_i \geq \epsilon$}
    {
        concatenate ($\mathcal{T}^{Shaved} \bigoplus \pi_i$)\;
    }
}

\caption{Shaving Hairs} \label{algo:shaving-hairs}
\end{algorithm}

\subsection{Straightening bumpy nodes}
\label{subsec:smoothing}

The adj-tree after the outrageous node removal step may still have some outrageous nodes, which need to be somehow fixed to improve the quality of the adj-tree to be the medial axis.
As a brute force removal of such a node may cause a disconnectedness of the adj-tree, it should be carefully managed.
This section presents an algorithm to remove those nodes outside $CA$ so that the adj-tree is a connected component.

Consider four consecutive nodes $n_1, n_2, n_3$, and $n_4$ of an adj-tree, and their associated three links $l_{12}, l_{23}$, and $l_{34}$.
Each link $l_{ij}$ is defined by two consecutive nodes $n_i$ and $n_j$.
Then we may define the unit direction vectors $\vec{u}_{12}, \vec{u}_{23}$, and $\vec{u}_{34}$ corresponding to $l_{12}, l_{23}$, and $l_{34}$, respectively.
Consider the difference vector $\vec{d}_{123} = \vec{u}_{23} - \vec{u}_{12}$.
Then $\| \vec{d}_{123} \|$ shows how far node $n_2$ is away from neighboring nodes.
A similar interpretation can be applied to $\| \vec{d}_{234} \|$ for node $n_3$.
Consider the following two conditions.

\begin{eqnarray}
\label{eqn:smoothing-condition-1}
\| \vec{d}_{123}                 \| > \alpha_1 \\
\label{eqn:smoothing-condition-2}
| \ \| \vec{d}_{234} \| - \| \vec{d}_{123} \| \ | > \alpha_2
\end{eqnarray}

\noindent where $\alpha_1$ and $\alpha_2$ are some threshold.
Satisfying both of the conditions means that node $n_2$ is relatively far away compared to node $n_3$
(Refer to~\ref{appendix:interpretation-of-condition-eqs} for an alternative interpretation of Eqs.~(\ref{eqn:smoothing-condition-1}) and (\ref{eqn:smoothing-condition-2})).
Then, it would be reasonable that the \emph{bumpy node} $n_2$ and both links $l_{12}$ and $l_{23}$ are removed from the adj-tree and a new link $l_{13}$ connecting $n_1$ and $n_3$ is inserted.
The proposed algorithm removes those nodes and associated links and inserts new links by checking the above conditions.
In this study, we have chosen 0.5 for both $\alpha_1$ and $\alpha_2$ through some experiments.
It turns out that the algorithm effectively smooths down by straightening the bumpy nodes (See Figs.~\ref{fig:V-graph-V-tree-shaved-trimmed-smoothed-trees-closeups} (c) and (d) to compare the adjacency tree before and after straightening bumpy nodes).

\section{Fused segmentation of ventricles and arteries}
\label{sec:segmentation-and-medical-applications}

This section presents the segmentation of the geometric mesh models $LV$ and $CA$ for left ventricle and coronary artery, respectively, using the above computed medial axis $\mathcal{M}$ of $CA$.
Once $\mathcal{M}$ is available, the segmentation can be done in a rather simple way.
We first present the segmentation of $LV$ in detail and then present that of $CA$, which is similar to but simpler than that of $LV$.

\subsection{Segmenting left ventricle}
\label{subsec:segmentation-of-left-ventricle-and-coronary-artery}

Given the medial axis $\mathcal{M}$, the algorithm will segment $LV$ into a set of subregions so that each subregion is assigned to one and only one node of $\mathcal{M}$.

Suppose that the constrained Delaunay triangulation $CDT^{LV}=(V, E, F, C)$ of $LV$ is available.
Let $N = \{ n_1, n_2, n_3, \ldots \}$ be the nodes of $\mathcal{M}$.
Then, we can formulate the segmentation of $LV$ as a problem to assign each cell of $C$ to one and only one node of $\mathcal{M}$ so that the summation of the distance between each cell and each node is minimized.
We use the Euclidean distance between the mass center of the cell and the coordinate of the medial axis node while other considerations such as the radii of CA, blood stream flow, and blood pressure may have to be reflected in future.

The segmentation of $LV$ is a many-to-one assignment problem that can be formulated as an integer linear program as follows:
\begin{eqnarray}
\label{eqn:ILP-obj-function}
Minimize &&  D(c_i, n_j) x_{ij} \\
\label{eqn:ILP-cons-1}
s.t. && \sum_{j \in J} x_{ij}   = 1, ~i~\in~I \\
\label{eqn:ILP-Variables}
&&   x_{ij} \in \{0, 1\}.
\end{eqnarray}
where $I$ and $J$ are the index sets which label the elements of $C$ and $N$, respectively.
$D(c_i, n_j)$ the distance between the mass center of $c_i$ and $n_j$.
Equation~(\ref{eqn:ILP-cons-1}) forces each cell to be assigned to one and only one node.
The formulation above can be solved by an implementation which, taking $O(|N||C|)$ time in the worst-case, assigns $c_i$ to the node which corresponds to
\begin{equation}
\label{eq:distance-function}
D^*(c_i) = Min_{j \in N} D(c_i, n_j).
\end{equation}
The definition of the distance in Eq.~(\ref{eq:distance-function}) can be improved by reflecting the medical condition, and the efficiency improvement needs to be made in the future.

\subsection{Segmenting the coronary artery}
\label{subsec:coronary-artery-segmentation}

The segmentation of $CA$ is similarly done by assigning each tetrahedral cell of the constrained Delaunay triangulation $CDT^{CA}$ of $CA$ to a node of $\mathcal{M}$ according to the minimum Euclidean distance.
Note that $CDT^{CA}$ is already available during the computation of $\mathcal{M}$.
Then the algorithm segments both left and right $CA$ by assigning each cell of $CDT^{CA}$s to a node of $\mathcal{M}$.

Figure~\ref{fig:CA-LV-segmentation} shows the segmentation result.
Figure~\ref{fig:CA-LV-segmentation}(a) shows the medial axis of both left coronary artery ($LCA$) and right coronary artery ($RCA$) is computed and their branches are also recognized:
The branches of $LCA$ are red-colored while those of $RCA$ are blue-colored.
Figures~\ref{fig:CA-LV-segmentation}(b) and (c) show the segmentations of $CA$ and $LV$ with the colors synchronized, respectively.
Figure~\ref{fig:CA-LV-segmentation}(d) shows the segmented $LV$ and $CA$ altogether from a different orientation.
Recall that we model left ventricle with a wall thickness: Be aware that the myocardium inside is also segmented and thus the volume of myocardial region corresponding to a supplying coronary artery piece can be measured.

\FigFour{CA-LV-segmentation}{0.32}{0.32}{0.32}{0.32}
{
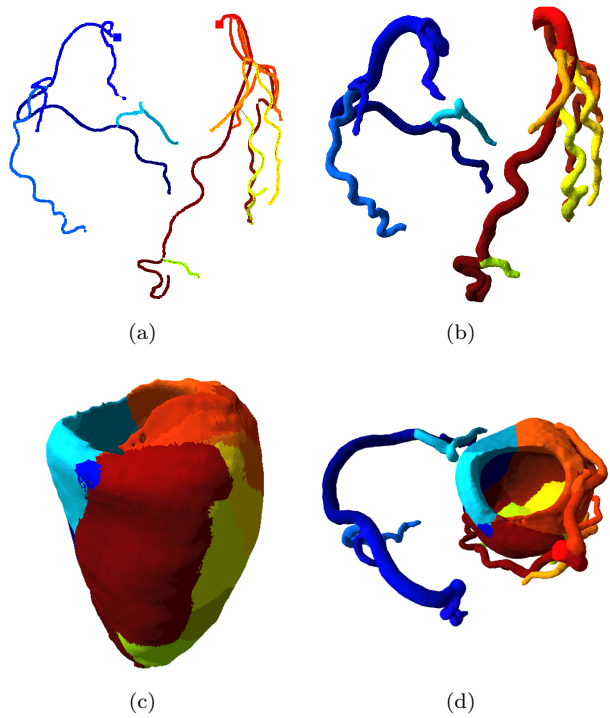
Segmentation of geometric models $LV$ and $CA$ for left ventricle and coronary artery, respectively, by using the medial axis of $CA$.
(a) The medial axis of both left coronary artery ($LCA$) and right coronary artery ($RCA$),
(b) the segmentation of $CA$ ($LCA$ and $RCA$),
(c) the segmentation of $LV$, and
(d) the segmented $LV$ and $CA$ altogether from a different orientation.
The colors of the segmented $LV$ and $CA$ are synchronized.
}

\section{Algorithm summary and experiments}
\label{sec:experiments-and-discussion}

This section summarizes the proposed algorithm and presents experimental results on both the segmentation of the coronary artery and left ventricle and computation of the medial axis.

\subsection{Algorithm summary}
The proposed algorithm is summarized in Algorithm~\ref{algo:segmenting-heart-via-approximate-medial-axis-of-coronoary-artery}, which first computes the constrained Delaunay triangulation $CDT^{CA}$ of $CA$.
Then an adjacency graph is constructed from $CDT^{CA}$ in Step 2.
Step 3 extracts an adjacency tree by removing the cycles of the adjacency graph.
The adjacency tree is transformed to the medial axis by removing outrageous nodes, shaving hairs, and straightening bumpy nodes in Steps 4, 5, and 6, respectively.
Step 7 computes $CDT^{LV}$ for the segmentation of $LV$.
Then the algorithm segments $LV$ and $CA$ into subregions by assigning each tetrahedral cell in both $CDT^{LV}$ and $CDT^{CA}$ to each node in the medial axis in Steps 8 and 9, respectively.

\begin{algorithm}[htpb]
\SetAlgoLined

\SetKwInOut{Input}{input}\SetKwInOut{Output}{output}
\Input{$CA$ and $LV$}
\Output{Medial axis $\mathcal{M}$ and segmentation of $CA$ and $LV$}
\BlankLine
Step 1) Compute constrained Delaunay triangulation $CDT^{CA}$ of $CA$\;
Step 2) Construct adjacency graph $\mathcal{G}$\ from $CDT^{CA}$\;
Step 3) Extract adjacency tree $\mathcal{T}$ from $\mathcal{G}$ (Algorithm~\ref{algo:construct-adjacency-tree})\;
Step 4) Remove the outrageous nodes of $\mathcal{T}$\;
Step 5) Shave hairs of $\mathcal{T}$ (Algorithm~\ref{algo:shaving-hairs})\;
Step 6) Straighten bumpy nodes of $\mathcal{T}$ (Eqs.~(\ref{eqn:smoothing-condition-1}) and (\ref{eqn:smoothing-condition-2}))\;
Step 7) Compute constrained Delaunay triangulation $CDT^{LV}$ of $LV$\;
Step 8) Segment $LV$\;
Step 9) Segment $CA$\;
\caption{Segmenting $LV$ and $CA$} \label{algo:segmenting-heart-via-approximate-medial-axis-of-coronoary-artery}
\end{algorithm}

Figure~\ref{fig:V-graph-V-tree-shaved-trimmed-smoothed-trees-closeups} shows the computation results for Steps 3, 4, 5 and 6 of Algorithm~\ref{algo:segmenting-heart-via-approximate-medial-axis-of-coronoary-artery} using the close-up of the adjacency tree and coronary artery.
Figures~\ref{fig:V-graph-V-tree-shaved-trimmed-smoothed-trees-closeups}(a) and (b) respectively show the adjacency trees before and after outrageous nodes are removed.
In both figures, the adjacency tree is superimposed on the boundary of the coronary artery mesh.
Figures~\ref{fig:V-graph-V-tree-shaved-trimmed-smoothed-trees-closeups}(c) and (d) show the adjacency trees only after the shaving and straightening, respectively.

\FigFour{V-graph-V-tree-shaved-trimmed-smoothed-trees-closeups}{0.4}{0.4}{0.4}{0.4}
{
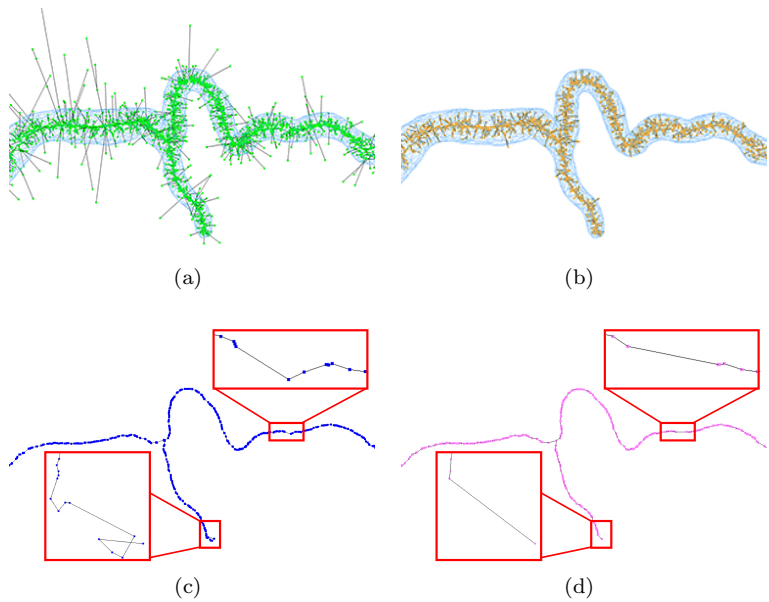
The computation results for Steps 3, 4, 5 and 6 of Algorithm~\ref{algo:segmenting-heart-via-approximate-medial-axis-of-coronoary-artery} (close-up of the adjacency tree (adj-tree) and coronary artery).
While the adj-tree is superimposed on the $CA$ mesh boundary for both (a) and (b), the adj-tree only is shown for both (c) and (d).
(a) The adj-tree of $CA$ before outrageous nodes removed,
(b) the adj-tree after outrageous nodes removed,
(c) the adj-tree after shaving hairs, and
(d) the adj-tree after straightening bumpy nodes.
}

\subsection{Experimental results}
\label{subsec:experimental_results}

Algorithm~\ref{algo:segmenting-heart-via-approximate-medial-axis-of-coronoary-artery} was implemented using the Microsoft Visual C++ and OpenGL library.
Figure~\ref{fig:VoroHeart} shows the developed \texttt{VoroHeart} program running on Microsoft Windows.
After segmenting $LV$ and $CA$ using $\mathcal{M}$, \texttt{VoroHeart} visualizes the recognized branches of $CA$ in the main pane with the synchronized colors for the corresponding $LV$ and $CA$ segments.
The right pane displays the tree hierarchy of $CA$ branches and the lower pane shows the parent-and-child relationship between $CA$ branches with a color-encoding.
The lower pane also shows the mass properties of each segmented subregion of $LV$ and its supplying $CA$ branch.
The mass properties include the volume/surface area of the segmented $LV$ subregion, the length, thickness, surface area, and volume of $CA$ branch, of which the importance for medical diagnosis and treatment to the cardiac function was verified~\cite{SumitsujiEtal16,KurataEtal15,FrangiEtal01}.
Hence, the \texttt{VoroHeart} program, which implemented the proposed algorithm, will be useful for the assessment of the severity of heart attack by quantifying the volume and area of the myocardium-at-risk.

\FigOne{VoroHeart}{0.9}
{
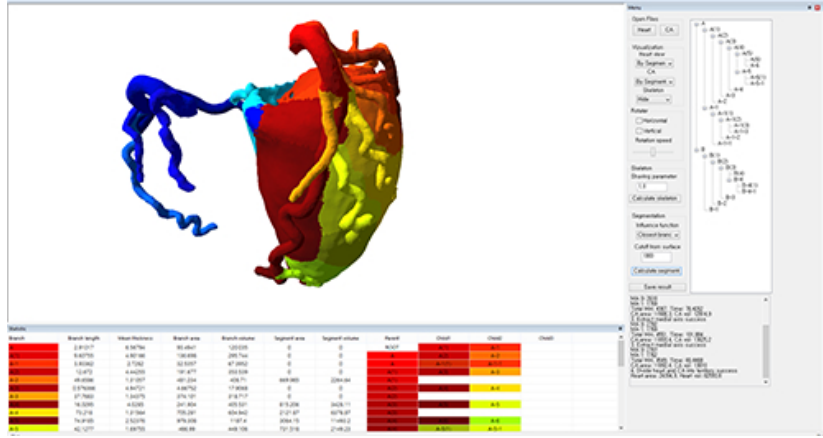
The developed \texttt{VoroHeart} program to segment the geometric models $LV$ and $CA$ of the left ventricle and coronary artery, respectively.
After segmenting $LV$ and $CA$ using $\mathcal{M}$, the \texttt{VoroHeart} program displays the recognized $CA$ branches and the corresponding segmented $LV$ subregions with the colors synchronized in the main pane.
The right pane shows the tree hierarchy of $CA$ branches.
The lower pane shows the mass properties of each segmented subregion of $LV$ and its supplying $CA$ branch.
The mass properties include the volume/surface area of the segmented $LV$ subregion, the length, thickness, surface area, and volume of the $CA$ branch
(Refer to the demo video for the details of the \texttt{VoroHeart}'s functions).
}

Figure~\ref{fig:CA-LV-closest-pair-correspondence} shows an application using segmentation of the left ventricle.
Suppose that we pick a point, as marked by the yellow arrow, which actually corresponds to a node of the medial axis of $CA$.
Consider that the coronary artery is obstructed at the picked point.
Figure~\ref{fig:CA-LV-closest-pair-correspondence}(a) shows
i) the subset of the coronary artery from the pick point down to the leaves (shown in light gray), and
ii) the subset of the myocardial muscle corresponding to the coronary artery subset (shown in dark gray).
Figure~\ref{fig:CA-LV-closest-pair-correspondence}(b) shows the case that the picking point is located further down to a leaf.
Observe that the corresponding myocardial muscle region shrinks.

\FigTwo{CA-LV-closest-pair-correspondence}{0.35}{0.35}
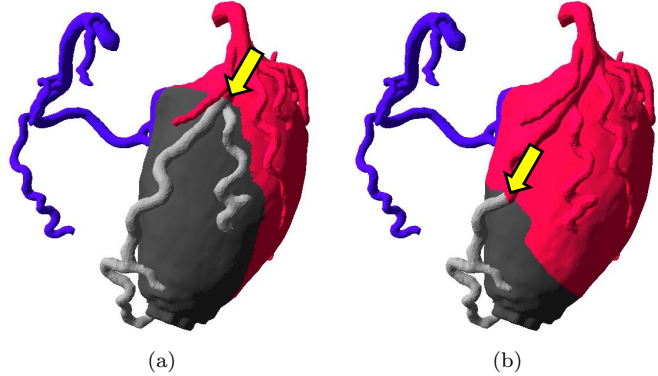
{Linkage between the geometric model $LV$ of left ventricle and the geometric model $CA$ of a supplying coronary artery.
Given the segmentation of $LV$ and $CA$, myocardium-at-risk can be precisely localized and quantified by designating the location of obstruction (yellow arrows).
(a) Obstruction in proximal $CA$ results in larger amounts of myocardium-at-risk compared to (b) obstruction in distal $CA$.
The extent and border of myocardium-at-risk can also be clearly identified.
The obstructive $CA$ pieces are shown in light grey and the subtended $LV$ pieces in dark grey.
}

Figure~\ref{fig:sectioning} shows a section view of the left ventricle and the entire coronary arteries.
Figure~\ref{fig:sectioning}(a) shows the remaining part of the left ventricle after the upper ventricular lump above a trimming plane is removed where the trimming plane (not shown in the figure) is created through a user-interaction with \texttt{VoroHeart} via the screen.
The figure in the red-box of Figure~\ref{fig:sectioning}(a) is the ventricle from a different view.
Be aware that the muscle in the ventricle wall is also properly assigned to a corresponding $CA$ branch.
Figure~\ref{fig:sectioning}(b) similarly shows a section view for a different trimming plane.
Thus we can investigate the morphometry of myocardium such as the wall thickness from the section view and can easily compute the thickness if necessary.
Note that the wall thickness of the left ventricle is an important measure for analyzing cardiac function and diagnosing cardiovascular disease~\cite{SasayamaEtal76,Gaasch79,OlivottoEtal03}.

Thus, the proposed geometric model based approach can facilitate various clinical studies where model quantification is an important measure~\cite{SaitoEtal05,SumitsujiEtal16,KurataEtal15,FrangiEtal01,PrakashEthier01}.
Furthermore, the proposed research could be exploited for applications related to model optimization.
For example, one of the promising therapies for cardiac disease is to transplant stem cells into either the myocardium at the site of injury or the supplying CA branch~\cite{SegersLee08,ShafiqEtal16}.
One important issue for this approach is to optimize the delivery of stem cells to the appropriate site so that cardiac regeneration is maximized \cite{Oettgen06}.
In this case, the segmentation result of this study would be more importantly used for the delivery optimization.

\FigTwo{sectioning}{0.5}{0.5}
{
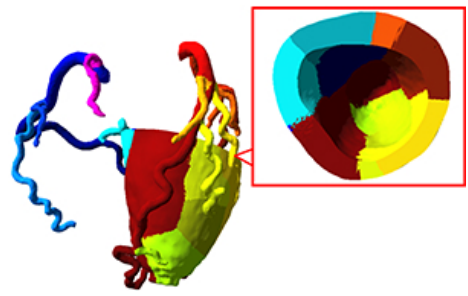
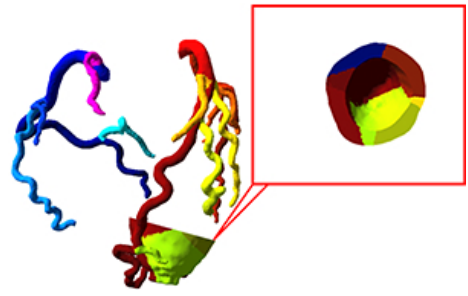
Section view of a left ventricle.
After the upper lump of the left ventricle above a trimming plane is removed, the remaining part of the left ventricle and the entire coronary arteries are shown in Figure~\ref{fig:sectioning}(a).
The trimming plane is created through a user-interaction with \texttt{VoroHeart} via screen (The trimming plane is not shown).
The figure in the red-box of Figure~\ref{fig:sectioning}(a) is the ventricle from a different view: Be aware that the muscle in the ventricle wall is also properly assigned to a corresponding $CA$ branch.
Figure~\ref{fig:sectioning}(b) similarly shows a section view for a different trimming plane.
}

We have also tested Algorithm~\ref{algo:segmenting-heart-via-approximate-medial-axis-of-coronoary-artery} using a data set of 20 clinical cases of anonymous patients from a teaching university hospital in Korea.
The cardiac CT image of each case for left and right coronary arteries and left ventricle was obtained using a dual source CT scanner, SOMATOM Definition Flash (Siemens Healthineers, Germany)~\cite{SiemensHealthineersHome} with a slice thickness 0.6mm and nonionic contrast medium (iomeron), in DICOM (Digital Imaging and Communications in Medicine) format~\cite{DICOMHome}.
The geometric model stored in the STL format was extracted from an individual cardiac CT image using the VitreaWorkstation program~\cite{VitreaHome}.
The computational environment for the experiment is as follows:
CUP: Inter Core2 Duo E7500 2.93Ghz; RAM: 4GB; OS: Windows 7.

Figure~\ref{fig:model_size} shows the size of the input geometric mesh models for both the coronary artery $CA$ (Figure~\ref{fig:model_size}(a)) and the left ventricle $LV$ (Figure~\ref{fig:model_size}(b)).
The horizontal axis, not only this figure but also throughout this section, represents the number of triangular faces of each input mesh model and the vertical axis the number of vertices and edges.
The graphs show a strong linear relationship.
Figures~\ref{fig:CDT_size}(a) and (b) show the size of $CDT$s for both $CA$ and $LV$, respectively.
The vertical axis represents the number of vertices, edges, faces, and cells of $CDT$s.
The graphs are again all linear.
Figure~\ref{fig:data_size_change}(a) shows the size of the adjacency graph and adjacency tree where different colors denote the entity size after different steps of the algorithm are applied.
Note that the data size decreases as the algorithm proceeds in its steps for computing the medial axis.
While the adjacency tree extraction and shaving steps significantly reduce the number of nodes, both the outrageous-node removing and the straightening steps do not reduce much.
Figure~\ref{fig:data_size_change}(b) shows two curves corresponding to the shaving and straightening steps.

\FigTwo{model_size}{0.47}{0.47}
{
Size of the input geometric mesh model for both
(a) the coronary artery and
(b) the left ventricle
}

\FigTwo{CDT_size}{0.47}{0.47}
{
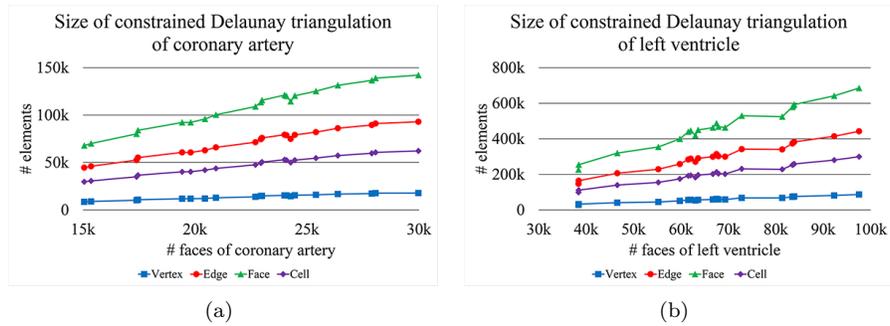
Size of the constrained Delaunay triangulation of the input geometric mesh model for both
(a) the coronary artery and
(b) the left ventricle
}

\FigTwo{data_size_change}{0.47}{0.47}
{
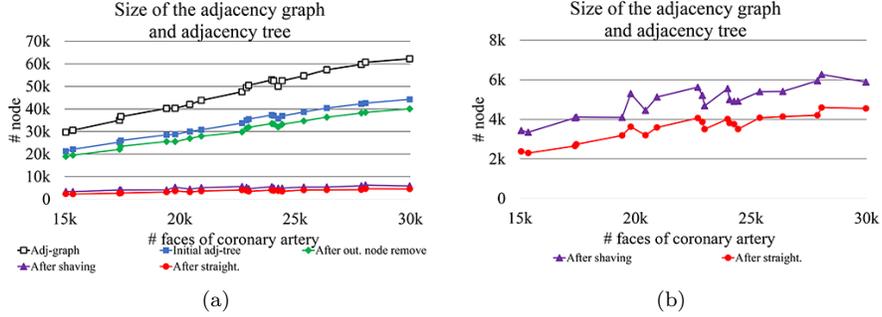
Size of the adjacency graph and adjacency tree: The data size decreases as the algorithm proceeds in its steps for computing the medial axis.
(a) Different colors denote the entity size after different steps of the algorithm are applied and
(b) only two curves corresponding to both shaving and straightening are shown.
}

Figure~\ref{fig:computation_time} shows the computation time of both the medial axis and the entire segmentation for the test data set of twenty clinical cases.
We group the steps of Algorithm~\ref{algo:segmenting-heart-via-approximate-medial-axis-of-coronoary-artery} into three phases.
Phase I (extraction of adjacency tree) consisting of Steps 1, 2, and 3;
Phase II (transformation of adjacency tree to medial axis) consisting of Steps 4, 5, and 6; and
Phase III (segmentation of ventricles and coronary arteries) consisting of Steps 7, 8, and 9.
Figure~\ref{fig:computation_time}(a) decomposes the computation time for the medial axis into three parts:
the time for loading each $CA$ model file,
the time for Phase I, and
the time for Phase II.
Note that the total time shows a quadratic increase with respect to the model size because the most expensive operation is Step 3 with the time complexity $O(|N||L|+{|N|}^2 log{|N|})$ with respect to $|N|$ nodes and $|L|$ links in the adjacency graph (See Sec.~\ref{subsec:transform-of-adjacency-graph-to-adjacency-tree}) and both $|N|$ and $|L|$ linearly increase regarding the model size.
For the clarity of the other times, the times for Phase I and II are shown in Figs.~\ref{fig:computation_time}(b) and (c), respectively.
The time for loading the $CA$ triangular mesh model is excluded.
Among the times for Phase I, the time for adjacency tree extraction is mostly dominant while times for both $CDT$ and the adjacency graph are relatively negligible.
For the steps of Phase II, the outrageous-node-removing takes more time than other steps.
The bumpy-node-straightening step is relatively negligible.
Figure~\ref{fig:computation_time}(d) shows the times for the $LV$ model file loading and Phase III, which consists of the computation of $CDT^{LV}$, $LV$-segmentation, and $CA$-segmentation.
The time for $CDT^{CA}$ is not shown because that is already included in Phase I.
The time for $LV$-segmentation shows a linear increase with model size
because its time complexity is $O(|N^\mathcal{M}||C|)$ with respect to $|N^\mathcal{M}|$ nodes in the medial axis $\mathcal{M}$ and $|C|$ tetrahedral cells in $CDT^{LV}$, respectively (See Sec.~\ref{sec:segmentation-and-medical-applications}).
Note that $|C| \gg |N^\mathcal{M}|$ and $|C|$ linearly increases with model size.
The similar argument applies to the time for $CA$-segmentation.

\FigFour{computation_time}{0.4}{0.4}{0.4}{0.4}
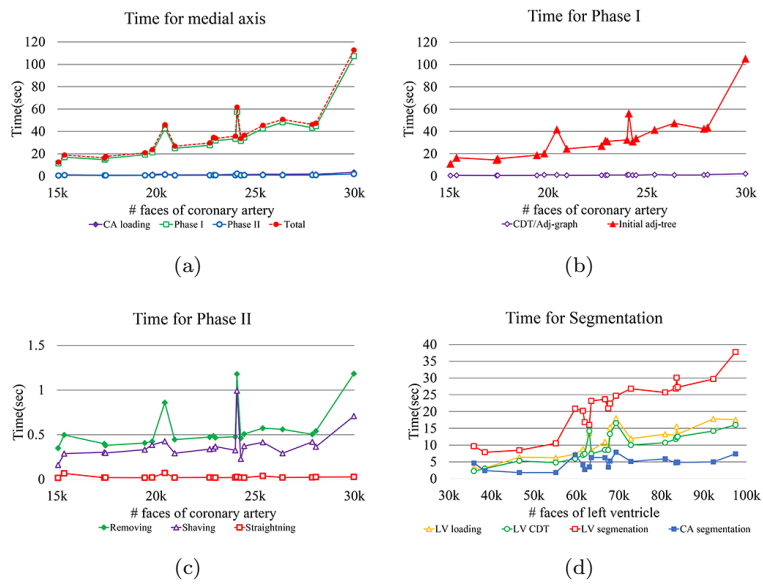
{Computation time for test data set.
(a) Decomposition of the medial axis computation time into $CA$ model file loading, Phase I, and Phase II,
(b) decomposition of times for Phase I into CDT/adj-graph and adj-tree,
(c) decomposition of times for Phase II into outrageous-node removing, hair shaving, and bumpy-node straightening, and
(d) times for $LV$ model file loading and Phase III, which is decomposed into $CDT^{LV}$ of $LV$, $LV$ segmentation, and $CA$ segmentation.}

\section{Conclusion}
\label{sec:conclusion}

This study presents an algorithm and its implementation to segment regional myocardium-at-risk subtended by any potentially obstructed coronary artery based on the geometric models of a triangular mesh for the coronary artery and myocardium obtained from an individual cardiac computed tomography image.
The key idea of the algorithm is
(i) computation of the medial axis of the coronary artery and
(ii) segmentation of the coronary artery and myocardium into a set of regions where each corresponds to a node of the medial axis.
The medial axis is transformed from an adjacency tree, which is extracted by removing cycles of an adjacency graph.
The adjacency graph is constructed from the constrained Delaunay triangulation of the triangular mesh model of the coronary artery.
The algorithmic accuracy and efficiency are theoretically asserted and experimentally verified.

Obstruction of the coronary artery results in acute myocardial infarction.
Hence, quantification of the regional amount of myocardium subtended by the obstructed coronary artery is of critical value in clinical medicine.
However, conventional methods such as the 17-piece model are inaccurate and frequently disagree with clinical practice.
The proposed algorithm provides a robust mathematical linkage between myocardium-at-risk and supplying coronary arteries so that ischemic myocardial region can be accurately identified, and both the extent and severity of myocardial ischemia can be quantified effectively and efficiently.
Furthermore, the computed result of segmented coronary artery and myocardium can be more importantly used for building optimization models of cardiac systems for various applications.
We believe that the algorithm and developed \texttt{VoroHeart} program will be an invaluable tool for patient-specific risk predictions and the treatment of obstructed coronary artery disease in clinical medicine.

\section*{Acknowledgments}

The authors would like to thank research assistants Seon-A Jeong and So-Hyeon Park for their efforts to prepare clinical data.
Jehyun Cha, Joonghyun Ryu, and Deok-Soo Kim were supported by the National Research Foundation of Korea(NRF) grant funded by the Korea government(MSIP) (No. 2017R1A3B1023591).
Jin-Ho Choi was supported by grants from Samsung Medical Center basic research (GL1B33211, CRP1500053), and the Korean Society of Interventional Cardiology (2014-1).

\section*{References}
\bibliography{biblio}

\appendix

\section{Proof of Lemma~\ref{lemma:monotonicity-of-distance-of-abstract-tree}}
\label{appendix:proof-1}
\begin{proof_}
Given $\mathcal{T}_{i-1}(N_{i-1},L_{i-1}), \mathcal{T}_i(N_i,L_i) \subseteq \mathcal{T}$ where $\mathcal{T}$ has $n$ leaf nodes,
$\Delta_{i-1} > \Delta_i$ because $\Delta_{i-1} = \Delta(\mathcal{T}_{i-1},\mathcal{T}) = \sum_{j=i}^n dist(n_j,\mathcal{T}_{i-1}) > \sum_{j=i+1}^n dist(n_j,\mathcal{T}_{i-1}) > \sum_{j=i+1}^n dist(n_j,\mathcal{T}_i) = \Delta_i$.

\end{proof_}

\section{ Interpretation of the conditions in Eqs.~(\ref{eqn:smoothing-condition-1}) and (\ref{eqn:smoothing-condition-2}) }
\label{appendix:interpretation-of-condition-eqs}

The conditions in Eqs.~(\ref{eqn:smoothing-condition-1}) and (\ref{eqn:smoothing-condition-2}) can be interpreted alternatively as follows.
Suppose that $n_1, n_2$, and $n_3$ are the uniformly sampled points on a smooth curve.
Thus, the lengths of $l_{12}$ and $l_{23}$ are equivalent to each other.
If we increase the sampling rate, the length of each link $l_{ij}$ decreases.
In the limiting case, each corresponding unit vector $\vec{u}_{ij}$ approaches the unit tangent vector of the curve and each link $l_{ij}$ approaches the curve piece.
Let $\vec{u}_{12}$ and $\vec{u}_{23}$ be the tangent vectors of the limiting case that correspond to the unit direction vectors of $l_{12}$ and $l_{23}$, respectively.
$\vec{d}_{123}$ is the difference vector on $n_2$ between $\vec{u}_{12}$ and $\vec{u}_{23}$.

Consider the osculating circle $\textbf{c}$ on the node $n_2$.
Let $o$ and $r$ be the center and the radius of $\textbf{c}$, respectively.
Let $\theta$ be the angle between $\vec{u}_{12}$ and $\vec{u}_{23}$ and
let $\mid l_{12} \mid = \mid l_{23} \mid = l$.
Then following Lemma~\ref{lemma:discrete-curvature} says that $\| \vec{d}_{123} \|$ can be represented via the radius $r$ of the osculating circle and the arc length $l$.

\begin{lemma_}
\label{lemma:discrete-curvature}
\begin{equation}
\label{eqn:discrete-curvature-1}
\| \vec{d}_{123} \| = \frac{ l } {r}
\end{equation}
\end{lemma_}

\begin{proof_}
Refer to Fig.~\ref{fig:discrete-curvature}.
The angle $\angle on_2n_1 = \frac{\pi - \theta}{2}$ because $\overline{on_2}$ is the angular bisector of $\angle n_1n_2n_3$.
Let us draw the perpendicular line $L_1$ from a point $m$ on $l_{12}$ to $\overline{n_2o}$.
Then $\sin{ \frac{\theta}{2} } = \frac{ \frac{\| \vec{d}_{123} \|}{2} }{\| \vec{u}_{12} \|}$ holds because $L_1$ is the angular bisector of $\angle n_2mm^{'}$ where $\angle n_2mm^{'} = \theta$.

Because the triangle $\triangle n_1on_2$ is an isosceles triangle, the angle $\angle n_1on_2 = \theta$.
Consider the perpendicular line $L_2$ from $o$ to $l_{12}$.
Then similarly, $\sin{ \frac{\theta}{2} } = \frac{ \frac{l}{2} }{r}$ holds because $L_2$ is the angular bisector of the angle $\angle n_1on_2$.
From the above equations, $\| \vec{d}_{123} \| = 2 \cdot \| \vec{u}_{12} \| \cdot \sin{ \frac{\theta}{2} } = 2 \cdot \frac{ \frac{l}{2} }{r} = \frac{ l } {r}$
\end{proof_}

\FigOne{discrete-curvature}{0.8}
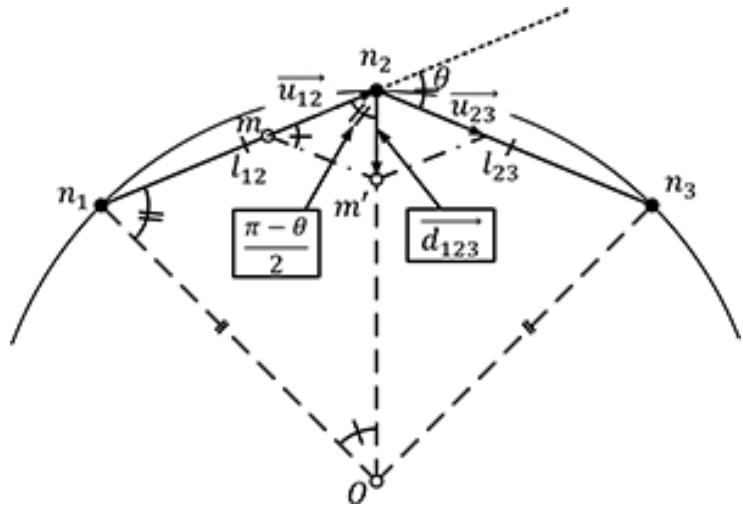
{Discrete curvature interpreted as the angle difference between tangent vectors}

Assuming that the length of each link is equal to each other and sufficiently small, i.e., $r \gg l$, Lemma~\ref{lemma:discrete-curvature} shows that $\| \vec{d}_{123} \|$ reflects the curvature $\frac{ 1 } {r}$ at node $n_2$ well.
The derivation and interpretation could similarly apply to $\| \vec{d}_{234} \|$ for $n_3$.
Equation~(\ref{eqn:smoothing-condition-2}) shows how much the curvature is different between $n_2$ and $n_3$.
Therefore, the proposed algorithm for straightening bumpy nodes approximately reflects the variation of the local curvature.

\end{document}